\documentstyle[amssymbols,fullpage]{article}

\newtheorem{prop}{{\bf Proposition}}

\newcommand{\pr}{{\bf Proof. }}

\newcommand{\beq}{\begin{equation}}
\newcommand{\eeq}{\end{equation}}
\newcommand{\bea}{\begin{eqnarray}}
\newcommand{\bean}{\begin{eqnarray*}}
\newcommand{\eea}{\end{eqnarray}}
\newcommand{\eean}{\end{eqnarray*}}
\newcommand{\demi}{\frac{1}{2}}

\newcommand{\sumi}{\sum_{i=1}^N}
\newcommand{\sumip}{\sum_{i=1}^p}
\newcommand{\sumjq}{\sum_{j=1}^q}
\newcommand{\sumij}{\sum_{_{\ i \neq j}^{i,j=1}}^N}
\newcommand{\la}{\lambda}
\newcommand{\Tr}{{\rm Tr}}
\newcommand{\w}{{\cal W}}
\newcommand{\RR}{R^{\pi}}
\newcommand{\poi}{\stackrel{\otimes}{,}}
\newcommand{\CC}{{\cal C}}
\newcommand{\un}{\hbox{\rm 1 \hskip -6pt l}}
\newcommand{\FF}{\tilde{F}}
\newcommand{\cR}{{\cal R}}
\newcommand{\cRR}{{\cal R}^{\pi}}
\newcommand{\WW}{\overline{W}}
\newcommand{\FFF}{\overline{F}}

\title{Observable Algebras for the Rational and Trigonometric \\
       Euler-Calogero-Moser Models}
\author{J. Avan $^*$ \and E. Billey \thanks{L.P.T.H.E. Universit\'e Paris
VI (CNRS UA 280), Box 126, Tour 16, $1^{{\rm er}}$ \'etage, 4 place Jussieu,
F-75252 PARIS CEDEX 05}}
\date{April 1994}

\begin{document}

\def\N{\Bbb N}
\def\Z{\Bbb Z}
\def\C{\Bbb C}
\def\R{\Bbb R}

\begin{titlepage}
\renewcommand{\thepage}{}
\maketitle
\vspace{2cm}
\begin{abstract}
We construct polynomial Poisson algebras of observables for the classical
Euler-Calogero-Moser (ECM) models. The conserved Hamiltonians and symmetry
algebras derived in a previous work are subsets of these algebras. We
define their linear, $N \rightarrow \infty$ limits, realizing $\w_{\infty}$
type algebras coupled to current algebras.
\end{abstract}

\vfill

PAR LPTHE 94-16

\end{titlepage}
\renewcommand{\thepage}{\arabic{page}}

\section{Introduction}
The Calogero-Moser (CM) models \cite{CMS} and their extensions to internal
degrees of freedom known as Euler-Calogero-Moser (ECM) models \cite{GHW} have
received a lot of attention in the past year. In particular, remarkable
algebraic structures were identified when considering specific sets of
invariant functions of the Lax matrices associated with these models. In
the case of CM models, polynomial Poisson-bracket algebras of
observables were constructed \cite{A} for rational and trigonometric
potentials; their $N \rightarrow \infty$ limit realized respectively the
classical linear $\w$ algebras $Sdiff(\R^2)$ and $Sdiff(\R
\times {\rm S}^1)$ \cite{A,HW1}. ECM models were shown to have
symmetry algebras: exact current algebras $\widehat{sl}(r)$ \cite{GHW} ---
in the rational case --- and exact Yangian algebras (quadratic deformations
of $\widehat{sl}(r)$) --- in the trigonometric case \cite{BAB1}. These
algebras were closely connected to the quantum algebra obtained for the
Haldane-type spin chain \cite{BGHP}.

More recently, it was suggested that a set of observables for the quantum
ECM models, realizing a $\w_{\infty}$ algebra when $N \rightarrow
\infty$, could be used to obtain the spectrum of the theory \cite{HW2}.
The finite $N$ case is however not so clear. This lead us to consider in
detail possible extensions of the previously known classical exact current
and Yangian algebras for finite $N$, using the techniques previously developed
to build the standard CM observable algebra \cite{A}. We shall here describe
general algebras of classical observables, associated both to the rational
and trigonometric ECM models; we shall compute the  Poisson-bracket
structures, and describe their $N \rightarrow \infty$ limit in connection
with the considerations in \cite{HW2}.

At this point, and before delving into the explicit construction of the
observable algebras, we need to state a keystone result:

\begin{prop}
\label{prop1}
Given a phase space ${\cal M}$ and a set of matrices $L_{(i)}$ on ${\cal M}$,
taking values in a Lia algebra ${\cal G}$ and realizing the Poisson bracket
structure:
\beq
  \label{pbs}
  \{ L_{(i)} \poi L_{(j)} \} =
  [ R, L_{(i)} \otimes \un] - [\RR, \un \otimes L_{(j)}]
\eeq
with $R \in {\cal G} \otimes {\cal G}$, the Poisson brackets of invariants
\beq
  \label{llpb}
  \left \{ \Tr \left ( L_{(1)}^{n_1} \cdots L_{(p)}^{n_p} \right ) ,
           \Tr \left ( L_{(1)}^{m_1} \cdots L_{(q)}^{m_q} \right )
  \right \}
\eeq
vanish exactly.
\end{prop}

\pr The expression (\ref{llpb}) takes the form:
\bean
  &   & \left \{ \Tr \left ( L_{(1)}^{n_1} \cdots L_{(p)}^{n_p} \right ) ,
                 \Tr \left ( L_{(1)}^{m_1} \cdots L_{(q)}^{m_q} \right )
         \right \}  \\
  & = &  \sumip \sumjq \sum_{a=1}^{n_i} \sum_{b=1}^{m_j} \Tr \left [
  L_{(1)}^{n_1} \cdots L_{(i-1)}^{n_{i-1}} L_{(i)}^{n_i-a} \otimes
  L_{(1)}^{m_1} \cdots L_{(j-1)}^{m_{j-1}} L_{(j)}^{m_j-b}
  \left ( [ R, L_{(i)} \otimes 1] - [\RR, 1 \otimes L_{(j)}] \right ) \right.
\\
  &   & \hspace{2.9cm} \left. L_{(i)}^{a-1} L_{(i+1)}^{n_{i+1}} \cdots
  L_{(p)}^{n_p} \otimes L_{(j)}^{b-1} L_{(j+1)}^{m_{j+1}} \cdots
  L_{(q)}^{m_q} \right ] \\
  & = & \sumip \sumjq \sum_{a=1}^{n_i} \sum_{b=1}^{m_j} \Tr \left [
  L_{(i)}^{a-1} L_{(i+1)}^{n_{i+1}} \cdots L_{(p)}^{n_p}
  L_{(1)}^{n_1} \cdots L_{(i-1)}^{n_{i-1}} L_{(i)}^{n_i-a} \otimes
  L_{(j)}^{b-1} L_{(j+1)}^{m_{j+1}} \cdots L_{(q)}^{m_q}
  L_{(1)}^{m_1} \cdots L_{(j-1)}^{m_{j-1}} L_{(j)}^{m_j-b} \right. \\
  &   & \hspace{2.75cm} \left. \left ( R \ L_{(i)} \otimes \un
                             - L_{(i)} \otimes \un \ R \right ) \right ]
  \hspace{1cm} + [ \RR-{\rm contribution} ]
\eean
by cyclicity of the trace. Eliminating the terms
$ - L_{(i)}^{a-1} (\cdots) L_{(i)}^{n_i-a}
  + L_{(i)}^{a} (\cdots) L_{(i)}^{n_i-a-1} $ when $a$ goes from $1$ to $n_i$
leaves us with only the extreme terms:
\bean
  &   & \sumip \sumjq \sum_{b=1}^{m_j} \Tr \left (
  L_{(i)}^{n_i} \cdots L_{(p)}^{n_p} L_{(1)}^{n_1} \cdots L_{(i-1)}^{n_{i-1}}
  \otimes L_{(j)}^{b-1} L_{(j+1)}^{m_{j+1}} \cdots L_{(q)}^{m_q}
  L_{(1)}^{m_1} \cdots L_{(j-1)}^{m_{j-1}} L_{(j)}^{m_j-b} R \right )  \\
  & - &  \sumip \sumjq \sum_{b=1}^{m_j} \Tr \left (
  L_{(i+1)}^{n_{i+1}} \cdots L_{(p)}^{n_p} L_{(1)}^{n_1} \cdots L_{(i)}^{n_i}
  \otimes L_{(j)}^{b-1} L_{(j+1)}^{m_{j+1}} \cdots L_{(q)}^{m_q}
  L_{(1)}^{m_1} \cdots L_{(j-1)}^{m_{j-1}} L_{(j)}^{m_j-b} R \right ) \\
  & + & [ \RR-{\rm contribution} ].
\eean
For fixed values of $j$ and $b$, the first-space terms cancel two
by two by obvious cyclicity of their form. Similarly the $\RR$ terms vanish
by the exact cyclic structure of the second-space terms in their
contribution.$\Box$

This result generalizes in a straightforward way the standard theorem for
a single Lax matrix with an $r$-matrix Poisson-structure \cite{S}. It can
be reexpressed in a form which will be more useful for us:

\begin{prop}
\label{prop2}
Given a Lax matrix $L$ and a set of dynamical matrices $A_{(i)}$ realizing
the Poisson algebra:
\bea
  \{ L \poi L \} & = & [R, L \otimes \un] - [\RR, \un \otimes L] \
  + {\rm extra \ terms} \nonumber \\
  \label{poisalg}
  \{ L \poi A_{(i)} \} & = & - [\RR, \un \otimes A_{(i)}] + \ {\rm extra \
  terms} \\
  \{ A_{(i)} \poi A_{(j)} \} & = & {\rm terms} \nonumber
\eea
the algebraic structure of the polynomial set $ \Tr ( L^n A_{(1)}^{m_1}
\cdots A_{(p)}^{m_p} ) $ is given solely by the extra terms in (\ref{poisalg}).
\end{prop}

\pr Apply prop.(\ref{prop1}) to the set $\{ L_{(i,\pm)} \} = \{ L \pm
A_{(i)} \}$ which realize the Poisson structure (\ref{pbs}) if $\{ L , A_{(i)}
\}$ realize (\ref{poisalg}) without extra terms and substitute $ A_{(i)} =
\demi ( L_{(i,+)} - L_{(i,-)} )$.$\Box$

\section{The rational Euler-Calogero-Moser model}
The system consists of $N$ particles on a line with pairwise $1/r^2$
interactions depending of their internal degrees of freedom. The phase space
is described by $2N$ conjugate coordinates $(p_i,q_i)_{i=1 \cdots N}$ and
$2 N r $ internal conjugate coordinates $(\eta_i^a,\xi_i^a)_{i=1 \cdots N,
a=1 \cdots r}$ parametrizing a coadjoint orbit in $sl(N)$ as $F_{ij}=
\sum_{a=1}^r \xi_i^a \eta_j^a.$ Integrability of the model requires to
restrict oneself to the submanifold $F_{ii}=\sum_{a=1}^r \xi_i^a \eta_i^a=
\alpha.$ The original Hamiltonian is:
\beq
  H = \demi \sumi p_i^2 - \demi \sumij \frac{F_{ij} F_{ji}}{(q_i-q_j)^2}.
\eeq
The equations of motion with spectral parameter $\la$ take the Lax form
$ \dot{L}(\la) = [ L(\la), M]$  with the Lax pair:
\bea
  L(\la) & = & L - \frac{1}{\la} F = \sumi p_i \ e_{ii}
  - \sumij \frac{F_{ij}}{q_i-q_j} \ e_{ij} - \frac{1}{\la}
  \sum_{i,j=1}^N F_{ij} \ e_{ij} \\
  M & = & \sumij  \frac{F_{ij}}{(q_i-q_j)^2} e_{ij}
\eea
Our algebra of observables will consist of traces of monomials
of the set
$$\{ L, F^{ab}=\sum_{i,j=1}^N \xi_i^b \eta_j^a \ e_{ij}, Q=\sumi q_i \
e_{ii} \}.$$ This choice is the generalization to the ECM model of
the observable algebras considered in the CM models \cite{A}. The choice
of invariant quantities as observables is natural since the ECM model, reduced
to orbits $(\xi_i^a,\eta_i^a)$, is a Hamiltonian reduction of a matrix-valued
model \cite{GHW,OP}, and the adjoint-invariant quantities are precisely those
which survive without modification the conjugation which redefines the
relevant reduced variables. These matrices realize the following Poisson
structure \cite{BAB2}:
\bea
  \label{basis}
  \{ L \poi L \} & = &  [ R, L \otimes \un] - [\RR, \un \otimes L]
                        - \sumij \frac{F_{ii}-F_{jj}}{(q_i-q_j)^2} \
                        e_{ij} \otimes e_{ji}  \nonumber \\
  \{ L \poi Q \} & = &  - [\RR, \un \otimes Q] + \CC \nonumber \\
  \{ L \poi F^{ab} \} & = &  - [\RR, \un \otimes F^{ab}] \\
  \{ Q \poi Q \} & = & \{ Q \poi F^{ab} \} = 0 \nonumber \\
  \{ F^{ab} \poi F^{cd} \} & = & ( \delta_{ad} F^{cb} \otimes \un
                                 - \delta_{cb} \un \otimes F^{ad} ) \ \CC
                                 \nonumber
\eea
where
$$ R = - \sumij \frac{1}{q_i-q_j} e_{ij} \otimes e_{ji} $$
and
$$ \CC = \sum_{i,j=1}^N e_{ij} \otimes e_{ji} $$
is the quadratic Casimir of the $sl(N)$ algebra.

We are therefore in the situation described by prop.(\ref{prop2}), guaranteeing
cancellation of the $R$-matrix contributions when computing the Poisson
structure of the monomial traces of $L$, $Q$, $F^{ab}$. Moreover, as usual,
we may also ignore the contribution $(F_{ii}-F_{jj})$ once we restrict the
system to the manifold $F_{ii}=\alpha$. Indeed, $F_{ii}$ generates the
conjugation of all matrices $L$, $F^{ab}$ and $Q$ (trivially)  by $U(1)$
acting on the $i^{\rm th}$ vector of the basis. Hence the Poisson brackets
of $F_{ii}$ with (adjoint-invariant) traces always vanish, and it is
consistent to compute the algebra of such quantities on the sub-manifold
$F_{ii}=\alpha.$

Furthermore, the generators $L$ and $Q$ obey the commutation relation:
\beq
  \label{comm}
  [ L , Q ] = F - \sumi F_{ii} e_{ii} = F - \alpha \un.
\eeq
We now describe the algebra of observables, using the following properties.

\begin{prop}
\label{prop3}
Given two matrices $A,B$, the monomial $\Tr ( A F^{ab} B F^{cd} ) $ is
rewritten as $\Tr ( A F^{cb} ) \Tr ( B F^{ad} ) .$
\end{prop}

It follows that we shall only retain monomials of order $0$ or $1$ in
$F^{ab}$. The proof of prop.(\ref{prop3}) is obvious, relying on the
projector structure of $F^{ab}$ as $F^{ab}_{ij}=\xi_i^b \eta_j^a.$

\begin{prop}
\label{prop4}
All monomials of the form $\Tr (A_{(1)} \cdots A_{(n)})$ and
$\Tr (A_{(1)} \cdots A_{(n)} F^{ab})$ with $A_{(i)} \in \{L,Q\}$ can be
written as polynomials of normal-ordered generators $\Tr (L^p Q^q)$ and
$\Tr (L^p Q^q F^{ab}), p,q \in \N.$
\end{prop}

\pr As in \cite{A} the commutation relation (\ref{comm}) and the projective
property (\ref{prop3}) allow a recursive proof of prop.(\ref{prop4}).
Specifically, denoting by $l$ the length of a monomial, i.e., the total
number of $L$ and $Q$ generators, one has:
\begin{itemize}
\item For $l=1$, normal-ordering is immediate.
\item If normal-ordering is ensured up to the length $l_0-1$, consider
first $\Tr (L \cdots Q)$ of length $l_0.$ If it is already normal-ordered,
the procedure stops; if not, normal-ordering is achieved by commuting
$Q$'s through $L$'s. Each such step eliminates two generators $L,Q$ and
creates one generator $F-\alpha \un$, thereby leaving residual terms of
length $l_0-2$ to which the recursion hypothesis applies. Note that, had
we allowed $Q^{-1}$ as a generator, the relevant commutation relation
$[L,Q^{-1}]=-Q^{-1} (F-\alpha \un) Q^{-1}$ prevents the normal-ordering
recursion. This will not occur in the trigonometric case. \\
Consider now $\Tr (L \cdots Q F^{ab})$. Each commutation operation eliminates
again two generators $L,Q$ and creates one generator $F-\alpha \un.$ The
projection property (\ref{prop3})  allows then factorization of the residual
terms into terms with one single $F^{cd}$ generator, of length at most
$l_0-2$, to which the recursion property then applies.$\Box$
\end{itemize}

Note that, as in \cite{A}, the normal-ordering of a given monomial may not
be unique; due to the particular form of the matrices and their finite size,
degeneracies will occur; in any explicit computation, they will be fixed
at every order by the choice of a reordering path for a given monomial.

\begin{prop}
\label{prop5}
The quantities $\Tr (L^p Q^q F)$  can be rewritten  as
polynomials of the variables $\Tr (L^n Q^m)$ and
$\Tr (L^n Q^m \FF^{ab})$ where $\FF^{ab}= F^{ab} - \delta_{ab}/r F$.
\end{prop}

\pr Again by recursion on the length $l=p+q$.
\begin{itemize}
\item When the length is 0 or 1,
\bean
  \Tr(F) & = & N \alpha \\
  \Tr(L F) & = & \Tr(L ([L,Q]+ \alpha \un)) = \alpha \Tr(L) \\
  \Tr(Q F) & = & \Tr(Q ([L,Q]+ \alpha \un)) = \alpha \Tr(Q).
\eean
\item Assuming that the proposition stands up to the length $l_0-1$, we take
$n+m=l_0.$ Then
$$ 0 = \Tr [L, L^n Q^m] =  m \Tr (L^n Q^{m-1} ( F - \alpha \un )) + \ {\rm
reordering} \  {\rm terms} \ {\rm of} \ L^n Q^{m-1} F.$$
Reordering terms of $L^n Q^{m-1} F$ contain two $F$'s and at most $l_0-2$
terms $L$ and $Q$. Hence they factorize into terms linear in $F^{ab}$,
following prop.(\ref{prop3}), and the factors normal-order following
prop.(\ref{prop4}). Finally every factor $\Tr (L^p Q^q {F}^{ab})$ is rewritten
as $ \Tr (L^p Q^q \FF^{ab}) + \delta_{ab}/r \Tr (L^p Q^q {F})$ with
$p+q \leq l_0-2$, to which the recursion hypothesis applies.$\Box$
\end{itemize}

It follows that the generators of our observable algebra to be considered
are reduced to the set:
\bea
  W_n^m & = & \Tr (L^n Q^m) \\
  \FF^{ab}_{n,m} & = & \Tr (L^n Q^m \FF^{ab}).
\eea

The Poisson algebra (\ref{basis}) is slightly modified to take into account
the change of generators:
\bea
 \label{modbasis}
 \{ \FF^{ab} \poi \FF^{cd} \} = &  &
    ( \delta_{ad} \FF^{cb} \otimes \un
    - \delta_{cb} \un \otimes \FF^{ad} ) \ \CC  \nonumber \\
    & - & \frac{\delta_{ab}}{r} ( \FF^{cd} \otimes \un - \un \otimes
    \FF^{cd} ) \ \CC - \frac{\delta_{cd}}{r} ( \FF^{ab} \otimes \un - \un
    \otimes \FF^{ab} ) \ \CC  \\
    & + & \frac{1}{r} ( \delta_{ad} \delta_{cb} - \frac{1}{r} \delta_{ab}
      \delta_{cd} ) (F \otimes \un - \un \otimes F ) \ \CC. \nonumber
\eea

The Poisson algebra of observables then follows from prop.(\ref{prop2}),
eq.(\ref{basis}) and (\ref{modbasis}) and prop.(\ref{prop3}) and (\ref{prop4}).

\bea
  \{ W_{n_1}^{m_1} , W_{n_2}^{m_2} \} = &   & \sum_{i=1}^{n_1}
  \sum_{j=1}^{m_2} \Tr (L^{n_1-i} Q^{m_1} L^{i-1} Q^{m_2-j} L^{n_2} Q^{j-1})
  \nonumber \\
  & - & \sum_{i=1}^{m_1} \sum_{j=1}^{n_2} \Tr (Q^{m_1-i} L^{n_1} Q^{i-1}
  L^{n_2-j} Q^{m_2} L^{j-1}) \nonumber  \\
  \label{wwr}
  = &   & (n_1 m_2 - n_2 m_1) W_{n_1+n_2-1}^{m_1+m_2-1} + {\rm ordering
  \ terms} \\
  \{ W_{n_1}^{m_1} , \FF_{n_2,m_2}^{ab} \} = &   &  \sum_{i=1}^{n_1}
  \sum_{j=1}^{m_2} \Tr (L^{n_2} Q^{j-1} L^{n_1-i} Q^{m_1} L^{i-1} Q^{m_2-j}
  \FF^{ab}) \nonumber \\
  & - & \sum_{i=1}^{m_1} \sum_{j=1}^{n_2} \Tr (L^{j-1} Q^{m_1-i} L^{n_1}
  Q^{i-1} L^{n_2-j} Q^{m_2} \FF^{ab}) \nonumber \\
  \label{wfr}
  = &   & (n_1 m_2 - n_2 m_1) \FF_{n_1+n_2-1,m_1+m_2-1}^{ab} + {\rm ordering \
  terms}
\eea

the ordering terms being of length $\leq n_1+n_2+m_1+m_2-4.$

\bea
  \label{ffr}
  \{ \FF_{n_1,m_1}^{ab} , \FF_{n_2,m_2}^{cd} \} = &   & \sum_{i=1}^{n_1}
  \sum_{j=1}^{m_2} \Tr (L^{n_1-i} Q^{m_1} \FF^{ab} L^{i-1} Q^{m_2-j} \FF^{cd}
  L^{n_2} Q^{j-1}) \nonumber \\
  & - & \sum_{i=1}^{m_1} \sum_{j=1}^{n_2} \Tr (Q^{m_1-i} \FF^{ab}
  L^{n_1} Q^{i-1} L^{n_2-j} Q^{m_2} \FF^{cd} L^{j-1}) \nonumber \\
  & + &  \delta_{ad} \ \Tr (L^{n_2} Q^{m_2} L^{n_1} Q^{m_1} \FF^{cb})
  - \delta_{cb} \ \Tr (L^{n_2} Q^{m_2} L^{n_1} Q^{m_1} \FF^{ad}) \nonumber \\
  & + & \frac{1}{r} \Tr \left \{ [ L^{n_1} Q^{m_1} , L^{n_2} Q^{m_2} ] \left (
  \delta_{ab} \FF^{cd} + \delta_{cd} \FF^{ab} - ( \delta_{ad} \delta_{cb}
  - \frac{1}{r} \delta_{ab} \delta_{cd} ) F \right ) \right \}.
\eea
The last term in (\ref{ffr}) generates a sub-leading contribution with
respect to the third term. The factorization property (\ref{prop3}) is now
replaced by:
\bea
\label{newfact}
  \Tr (A \FF^{ab} B \FF^{cd}) = &   & \Tr (A F^{cb}) \Tr (B F^{ad})
  - \frac{\delta_{ab}}{r} \sum_{i=1}^{r} \Tr (A F^{ci}) \Tr (B F^{id})
  - \frac{\delta_{cd}}{r} \sum_{i=1}^{r} \Tr (A F^{ib}) \Tr (B F^{ai})
  \nonumber \\
  & + & \frac{\delta_{ab} \delta_{cd}}{r^2} \sum_{i,j=1}^{r}
  \Tr (A F^{ij}) \Tr (B F^{ji})
\eea
which is then converted into a canonical expression in terms of
$\Tr (A \FF^{ab})$ by using prop.(\ref{prop5}).  From prop.(\ref{prop5})
we know that normal-order reexpresses $\Tr(L^nQ^mF)$ as $\alpha \Tr(L^nQ^m)
+ \ {\rm lower \ order \ polynomials}.$ Hence, keeping only the explicit
highest order of each term in (\ref{ffr}) we get:
\bea
\label{ffr2}
  \{ \FF_{n_1,m_1}^{ab} , \FF_{n_2,m_2}^{cd} \} = &   & \sum_{i=1}^{n_1}
  \sum_{j=1}^{m_2} \FF^{cb}_{n_1+n_2-i,m_1+j}  \FF^{ad}_{i-1,m_2-j-1}
  + S_1 \nonumber \\
  & - & \sum_{i=1}^{m_1} \sum_{j=1}^{n_2} \FF^{cb}_{i,m_1-j}
  \FF^{ad}_{n_1+n_2-i,m_2+j} + S_2  \nonumber \\
  & + & \delta_{ad} \FF^{cb}_{n_1+n_2,m_1+m_2}
  - \delta_{cb} \FF^{ad}_{n_1+n_2,m_1+m_2} + S_3 \nonumber \\
  & + & S_4.
\eea
$S_1$ and $S_2$ contain the extra terms in (\ref{newfact}) due to the
redefinition of $\FF$, plus ordering terms. $S_3$ contains reordering terms,
and $S_4$ contains the purely reordering terms arising from the commutator-
like term in (\ref{ffr}), consequence of the Poisson-bracket modification
(\ref{modbasis}).

\bigskip

At this point a number of remarks are to be made:
\begin{description}
\item[{\bf 1. No reordering terms.}] A number of Poisson-brackets (\ref{wwr},
\ref{wfr},\ref{ffr2}) have no extra reordering terms. This is true each time
the second term in the Poisson-bracket expression contains only one generator
$L,Q$ at any power and another generator $L,Q,\FF^{ab}$ at power 1. One has:
\beq
  \begin{array}{lcll}
      \{  W_n^0 , W_m^0 \} & = & 0 & ( N \ {\rm first \
      commuting \ ECM \  Hamiltonians \ \cite{GHW,BAB2}} ) \\
      \{  W_n^0 , W_m^1 \} & = &  m W_{n+m-1}^0 & ( W_m^1 \ {\rm
      as \ ``intertwiner" \ between \ Hamiltonians} ) \\
      \{  W_n^1 , W_m^1 \} & = &  (n-m) W_{n+m-1}^1 & ( {\rm
      centerless \ Virasoro \ algebra, \ mentioned \ in \ \cite{HW2} ).}
  \end{array}
\eeq
This closed set generates a symmetric Lie algebra. Note that, as a curiosity,
one also has:
\beq
  \{  W_n^2 , W_m^0 \} = 2 m W_{n+m-1}^1.
\eeq
The current algebra symmetry also belongs to this class \cite{GHW,BAB1}:
\beq
  \begin{array}{rcl}
    \{  W_n^0 , F_{m,0}^{ab} \} & = & 0 \\
    \{  F_{m 0}^{ab} , F_{n,0}^{cd} \} & = & \delta_{ad} \ F^{cb}_{n+m,0}
                                              - \delta_{cb} \ F^{ad}_{n+m,0}.
  \end{array}
\eeq
The Virasoro algebra acts on the current algebra as:
\beq
  \{  W_n^1 , F_{m,0}^{ab} \} = - m F^{ab}_{n+m-1,0}.
\eeq
No extra reordering terms also occurs in a different class of Poisson
brackets:
\beq
  \label{step1hw}
  \{  W_n^0 , W_0^m \} = n m W_{n-1}^{m-1}.
\eeq
This structure is related, as we shall see, to the classical version of the
construction of the observables in \cite{HW2}.
\item[{\bf 2. Different choices of basis.}] Absence of normal-ordering
con\-tri\-bu\-tions also oc\-cur when com\-put\-ing Pois\-son-brackets of
very specific polynomials of the original generators ${W,\FF}$; they correspond
in fact to relevant non-linear changes of basis: the first typical example
consists of the higher Hamiltonians. Only the $N$ first Hamiltonians are
simple generators $W_n^0$; the higher ones are polynomials in $L$ and $\FF$,
typically Weyl-ordered monomials of $L^n$ and $F^m$
$$ H_n^m = \oint \frac{d \la}{2 {\rm i} \pi \la^{m+1}} ( L + \la F ) ^{n+m}. $$
Since $F=\sum_{a=1}^r F^{aa}$,  the conserved higher Hamiltonians are
particular polynomials in $\Tr (L^n \FF^{ab})$, necessarily scalar under
$sl(r)$, and their commutation follows much more easily from direct
computation using the spectral-parameter Lax operator \cite{BAB2}.  \\
Another set of exactly Poisson-commuting Hamiltonians can be obtained by direct
construction, following the procedure in \cite{A}. We define:
\beq
  L^{\pm} = L \pm Q \pm \omega  Q^2  \hspace{2cm} ( \omega \in \R ).
\eeq
 From the Poisson structure (\ref{basis}), and repeating the derivation in
\cite{A} one gets:
\bea
  \{ L^+ L^- \poi L^+ L^- \} = &   & [ \cR , L^+ L^- \otimes \un ]
  - [ \cRR , \un \otimes L^+ L^- ]  \nonumber \\
  & - & 2 \ \CC \ \left ( L^+ L^- \otimes \un - \un \otimes L^+ L^- \right ) \\
  & - & 2 \ \omega \ \CC \ \left ( L^+ Q L^- \otimes \un - \un \otimes
  L^+ Q L^- + L^+ L^- \otimes Q - Q \otimes L^+ L^- \right ) \nonumber
\eea
with
\beq
  \cR = (\un \otimes L^+ ) \ R + R \ (\un \otimes L^-).
\eeq
Although not an exact $r$-matrix structure, this nevertheless guarantees the
commutation of the $N$ first Hamiltonians $ H^{(n)} = 1/2n \ \Tr (L^+ L^-)^n.$
In particular this set includes a natural one-body extension of the
rational ECM Hamiltonian:
\beq
  H^{(1)} = \demi \sumi p_i^2 - \demi \sumij \frac{F_{ij} F_{ji}}{(q_i-q_j)^2}
            - \demi \sumi ( q_i^2 + 2 \omega q_i^3 + \omega^2 q_i^4).
\eeq
This extension was indeed considered in \cite{Po} and a set of $N$ commuting
Hamiltonians was constructed. Note that full integrability would require
constructing an extra set of $N(r-1)$ commuting Hamiltonians, which we do not
see clearly how to get at the moment. \\
The quantities $\Tr (L^+ L^-)^n$ naturally belong to our algebra of
observables. They are however not naturally normal-ordered in terms of $L$
and $Q$, hence they are expressed as complicated polynomials of our generators
--- except $H^{(1)} = 1/2 \ \Tr (L^2+Q^2+2\omega Q^3+\omega^2 Q^4 ).$
Obviously the associated algebra of observables from which
to deduce exact eigenfunctions of the Hamiltonian flows should be extracted
from  the $\{L^+,L^-\}$ representation. This represents a non-linear change of
variables; but the identical normal-ordering procedure (in terms of $L^{\pm}$)
and algebra structure will be obtained in a similar way to the case of
${L,Q}$ since the Poisson structure and commutation relations are essentially
of the same form. Suitability of the basis depends on which problem ---
precisely which Hamiltonians--- one considers. A similar connection relates
the algebras obtained in \cite{A} and \cite{HW1} for the CM model.
\item[{\bf 3. The Hikami-Wadati algebra.}] The algebra considered in \cite{HW2}
consisted of generators obtained by successively commuting the quantum
Hamiltonians by the operator $\Tr(Q^2)$. These quantum Hamiltonians are in
fact obtained from expanding the quantum determinant of the generating
transfer matrix \cite{BGHP}. It is to be expected that their classical limit
is equivalently described by the Poisson-commuting classical traces
$\Tr(L^n)$. Hence the classical construction of the Hikami-Wadati algebra
is achieved by bracketing $\Tr(L^n)$ by $\Tr(Q^2)$, the first step of which
is one of the Poisson-brackets (\ref{step1hw}). This is easily seen to
substitute at each step one term $p_i$ into one term $q_i$, hence leading to
a set of the form:
\beq
  \label{weylgen}
  W^{n,(s)}_{HW} = \frac{1}{C^{s-1}_{n+2(s-1)}} \oint
  \frac{d\omega}{2 {\rm i} \pi \omega^s} \Tr(L+\omega Q)^{n+2(s-1)}.
\eeq
This follows from the fact that for any matrices $A,B$,
$$ \{ \Tr(A L B), \Tr(Q^2) \} = \Tr(A Q B) + \ {\rm contributions \ from \ }
   A,B. $$
Hence the $W^{n,(s)}_{HW}$ are a set of ``Weyl-ordered'' generators. However,
for $N$ finite, they do not close an algebra. Indeed
$$ \{ W^{1,(2)}_{HW} , W^{1,(3)}_{HW} \} = W^{2,(3)}_{HW} + \frac{3}{10}
   [ - 5 \alpha^2 W^{2,(1)}_{HW} + 3 \sum_{a,b=1}^r \Tr(L^2 F^{ab})
   \Tr(F^{ba}) + 2 \sum_{a,b=1}^r \Tr(L F^{ab}) \Tr(L F^{ba})]. $$
The last quantity is a typical reordering contribution, not to be obtained
from the explicitely symmetric Weyl-ordered generators (\ref{weylgen}). One
can compute its $p_i p_j \ (i \neq j)$ term which is $2 F_{ij} F_{ji}$: it
is non zero, dynamical and does not contain $1/(q_i-q_j)$, hence it lies
outside the original algebra $W^{n,(s)}_{HW}$.
\end{description}

This third remark deserves elaboration. It is nevertheless possible to define
a large $N$ limit of the observable algebra in which the two sets $W_n^m$
and $W^{n,(s)}_{HW}$ are identified and then close a Poisson algebra.
Redefining
\bea
  W_n^m & = & N^{n+m-2} \ \WW_n^m \\
  \FF^{ab}_{n,m} & = & N^{n+m} \ \FFF^{ab}_{n,m}
\eea
eliminates all reordering terms when $N \rightarrow \infty$, since reordering
always decreases the number $n+m$ of $L$ and $Q$ terms ---not forgetting that
the term $\Tr (L \cdots Q  \cdots F)$, which arises as first step in
reordering of  $W_n^m$ and would be of order $N^{n+m-2}$ also, is in fact
reexpressed through prop.(\ref{prop5}) as polynomials of $(L,Q)$ of length
$n+m-4$, hence is actually of lower order.

In this limit:
\bea
  \{ \WW_{n_1}^{m_1} , \WW_{n_2}^{m_2} \} & = & (n_1 m_2 - n_2 m_1)
     \WW_{n_1+n_2-1}^{m_1+m_2-1} \\
  \{ \WW_{n_1}^{m_1} , \FFF^{ab}_{n_2,m_2} \} & = & (n_1 m_2 - n_2 m_1)
     \FFF^{ab}_{n_1+n_2-1,m_1+m_2-1} \\
  \{ \FFF^{ab}_{n_1,m_1} , \FFF^{cd}_{n_2,m_2} \} & = & \delta_{ad} \
  \FFF^{cb}_{n_1+n_2,m_1+m_2} - \delta_{cb} \FFF^{ad}_{n_1+n_2,m_1+m_2}.
\eea
Indeed the first algebra realizes the structure of ${\rm Sdiff}(\R^2) \sim
{\cal W}_{\infty}$, classical limit of the quantum algebra defined in
\cite{HW2}; the $N \rightarrow \infty$ limit achieves the decoupling of the
two algebras ${\cal W}_{\infty}$ and $sl(r) \otimes \C(\la) \otimes \C(\mu)$
(double loop algebra of $\FF^{ab}_{n m}$).

\section{The trigonometric Euler-Calogero-Moser model}
The phase space structure is the same as before. The Hamiltonian is now:
\beq
  H = \demi \sumi p_i^2 - \demi \sumij \frac{F_{ij} F_{ji}}{\sinh^2(q_i-q_j)}.
\eeq
The Lax matrix with spectral parameter takes the form:
\beq
  L(\la) = L - (\coth(\la)+1) F
\eeq
where
\bea
  F & = & \sum_{a=1}^r \sum_{i,j=1}^N \xi_i^a \eta_j^a \ e_{ij} \\
  L & = & \sumi (p_i+F_{ii}) \ e_{ii} + 2 \sumij \frac{F_{ij}}
 {1-{\rm e}^{2(q_i-q_j)}} \ e_{ij}.
\eea
The Hamiltonian is given by $H=1/2 \ [ \Tr(L^2) - 2 \alpha p - N \alpha^2 ].$

As in the rational case, and for the same reason, a natural choice of
observables is the set of (ad-invariant) traces of monomials of the generators
$ \{ L, F^{ab},K=\exp(2Q),K^{-1} \}$. They realize the following Poisson
structure \cite{BAB1}:
\bea
\label{basistr}
  \{ L \poi L \} & = & [R,L \otimes \un ] - [\RR,\un \otimes L ]
                       + \sumij \frac{F_{ii}-F_{jj}}{\sinh^2(q_i-q_j)}
                      \ e_{ij} \otimes e_{ji} \\
  \{ L \poi K^{\pm1} \} & = & - [\RR,\un \otimes K^{\pm1} ]
                              \pm ( K^{\pm1} \otimes \un
                                   + \un \otimes K^{\pm1} ) \ \CC  \\
  \{ L \poi F^{ab} \} & = & - [\RR,\un \otimes F^{ab} ]
                            + [ \CC, F^{ab}] \\
  \{ K^{\pm1} \poi K^{\pm1} \} & = & \{ K \poi K^{-1} \} =
  \{ K^{\pm1} \poi F^{ab} \} = 0 \\
  \{ F^{ab} \poi F^{cd} \} & = & ( \delta_{ad} F^{cb} \otimes \un
                                 - \delta_{cb} \un \otimes F^{ad} ) \ \CC
\eea
where
$$ R = - \sumij \coth(q_i-q_j) \ e_{ij} \otimes e_{ji},$$
$\CC$ still being the Casimir element of ${\rm sl}(N).$

The commutation relations involved in the normal-ordering procedure
become (with $F_{ii}=\alpha)$:
\bea
  \label{comk}
  [L,K] & = & 2 F K - 2 \alpha K \\
  \label{comk-1}
  [L,K^{-1}] & = & -2 K^{-1} F + 2 \alpha K^{-1}.
\eea
The algebra of relevant, algebraically independant observables, follows from
prop.(\ref{prop3}) and the two  equivalents of prop.(\ref{prop4}) and
(\ref{prop5}) describing normal-ordering. Here however both positive and
negative powers of $K$ are allowed.

\begin{prop}
\label{prop6}
All monomials $\Tr(A^{(1)} \cdots A^{(n)})$ and $\Tr(A^{(1)} \cdots
A^{(n)} F^{ab})$ with $A \in \{ L,K,K^{-1} \}$ can be rewritten as polynomials
of normal-ordered generators $\Tr(L^p K^q)$ and $\Tr(L^p K^q F^{cd}) \equiv
F^{cd}_{p,q}, \ p \in \N, q \in \Z.$
\end{prop}

\pr similar as prop.(\ref{prop4}). From relations (\ref{comk}) and
(\ref{comk-1}), commuting $L$ through $K$ or $K^{-1}$ to normal-order
decreases by 1 the number of $L$ terms and creates an $F$ term. Hence a
recursion proof on the total number of $L$ terms is derived
straightforwardly.$\Box$

\begin{prop}
\label{prop7}
Monomials of the form $\Tr(L^n K^mF)$ with $m \neq 0$ can be rewritten as
polynomials of the variables $\Tr(L^p K^q)$ and
$\Tr(L^p K^q \FF^{ab}).$
\end{prop}

\pr As in prop.(\ref{prop5}), the commutation relations  (\ref{comk},
\ref{comk-1}) allow to define a recursive procedure.
\begin{itemize}
\item When the length (i.e., the number of $L$ terms) is 0,
\bean
  \Tr(K F) & = & \demi \Tr([L,K]+2 \alpha K) = \alpha \Tr(K) \\
  \Tr(K^{-1} F) & = & - \demi \Tr([L,K^{-1}]-2 \alpha K^{-1})
                     = \alpha \Tr(K^{-1}).
\eean
\item We assume the proposition to be true up to the length $n-1$. One has
for $m \neq 0$:
$$ 0 = \Tr [ L,L^n K^m] = 2 m \Tr(L^n Km F ) - 2 \alpha m \Tr(L^n K^m ) +
{\ \rm ordering \ terms,}$$
with ordering terms of length $n-1$ which are then written using prop.(
\ref{prop3}) and recursion hypothesis. This proof collapses when $m=0$.$\Box$
\end{itemize}

Hence our generators are chosen to be
\bea
  W_n^m & = & \Tr(L^n K^m) \nonumber \\
  \label{observ}
  F_n^{ab} & = & \Tr(L^n F^{ab}) = F^{ab}_{n,o} \\
  \FF_{n,m}^{ab} & = & \Tr(L^n K^m \FF^{ab}) \hspace{2cm}{\rm with \ }
  m \neq 0. \nonumber
\eea
The Poisson algebra of observables now follows from (\ref{basistr}) and
(\ref{observ}):
\bea
  \{ W_{n_1}^{m_1} , W_{n_2}^{m_2} \} = &   & \sum_{i=1}^{n_1}
  \sum_{j=1}^{m_2} \Tr (L^{n_1-i} K^{m_1} L^{i-1} ( K^{m_2-j+1} L^{n_2}
  K^{j-1} + K^{m_2-j} L^{n_2} K^j )) \nonumber \\
  & - & \sum_{i=1}^{m_1} \sum_{j=1}^{n_2} \Tr ((K^{m_1-i} L^{n_1} K^{i}
  + K^{m_1-i+1} L^{n_1} K^{i-1} ) L^{n_2-j} K^{m_2} L^{j-1}) \nonumber  \\
  \label{wwtr}
  = &   & 2 (n_1 m_2 - n_2 m_1) W_{n_1+n_2-1}^{m_1+m_2}
  + {\rm ordering \ terms}
\eea
\bea
  \{ W_{n_1}^{m_1} , F_{n_2,m_2}^{ab} \} = &   &  \sum_{i=1}^{n_1}
  \sum_{j=1}^{m_2} \Tr ( L^{n_1-i} K^{m_1} L^{i-1} ( K^{m_2-j+1} F^{ab}
  L^{n_2} K^{j-1} +  K^{m_2-j} F^{ab} L^{n_2} K^{j} )) \nonumber \\
  & - & \sum_{i=1}^{m_1} \sum_{j=1}^{n_2} \Tr (( K^{m_1-i} L^{n_1}
  K^{i} + K^{m_1-i+1} L^{n_1} K^{i-1} ) L^{n_2-j} K^{m_2} F^{ab} L^{j-1})
  \nonumber \\
  & + & \sum_{i=1}^{n_1} \Tr([L^{n_2} K^{m_2},L^{n_1-i} K^{m_1} L^{i-1}]
  F^{ab}) \nonumber \\
  \label{wftr}
  = &   & 2 (n_1 m_2 - n_2 m_1) F_{n_1+n_2-1,m_1+m_2}^{ab} + {\rm ordering \
  terms}
\eea
\bea
  \{ F_{n_1,m_1}^{ab} , F_{n_2,m_2}^{cd} \} = &   & \sum_{i=1}^{n_1}
  \sum_{j=1}^{m_2} \Tr(L^{n_1-i} K^{m_1} F^{ab} L^{i-1}
  ( K^{m_2-j+1} F^{cd} L^{n_2} K^{j-1} + K^{m_2-j} F^{cd} L^{n_2} K^{j} ))
  \nonumber \\
  & - & \sum_{i=1}^{m_1} \sum_{j=1}^{n_2} \Tr (( K^{m_1-i} F^{ab} L^{n_1}
  K^i + K^{m_1-i+1} F^{ab} L^{n_1} K^{i-1} ) L^{n_2-j} K^{m_2} F^{cd} L^{j-1} )
  \nonumber \\
  & + & \sum_{i=1}^{n_1} \Tr(L^{n_1-i} K^{m_1} F^{ab} L^{i-1} [ F^{cd},L^{n_2}
  K^{m_2} ] ) \nonumber \\
  & - & \sum_{i=1}^{n_2} \Tr ( [ F^{ab},L^{n_1} K^{m_1} ] L^{n_2-i} K^{m_2}
  F^{cd}   L^{i-1} ) \nonumber \\
  \label{fftr}
  & + & \delta_{ad} \Tr(L^{n_2} K^{m_2} L^{n_1} K^{m_1} F^{cb} )
  - \delta_{cb} \Tr(L^{n_1} K^{m_1} L^{n_2} K^{m_2} F^{ad} ).
\eea
For the sake of simplicity we have expressed the last Poisson bracket
(\ref{fftr}) in terms of unreduced generators $F^{ab}_{n,m}$ instead of
$\FF^{ab}_{n,m}$ for $m \neq 0$. When expressed in terms of the exact
generators $\FF_{n,m}^{ab}$, expressions of analogous form to
(\ref{ffr},\ref{newfact}) arise. (\ref{wftr}) is not modified and (\ref{fftr})
acquires extra terms of the form (\ref{newfact}).
Again, some Poisson brackets do not generate
reordering terms; this is the case whenever only one term ($L$ or $K$) is
present in the monomial. In particular:
\bea
  \{ W_n^0 , W_m^0 \} & = & 0 \hspace{1cm} { \rm : \ commuting \
                                                     Hamiltonians} \\
  \{ W_n^0 , W_m^1 \} & = & 2 n W_{n+m-1}^1 \\
  \{ W_n^0 , W_0^m \} & = & 2 n m  W_{n-1}^m
\eea
but $ \{ W_n^1 \poi W_m^1 \} = W_{n+m-2}^2 + \ {\rm ordering \ terms}.$

As previously derived, the Yangian algebra is also exact \cite{BAB1}:
\bea
  \{ W_n^0 , F^{ab}_{m,0} \} & = & 0 \nonumber \\
  \{ F^{ab}_{n,0} , F^{cd}_{m,0} \} & = & \delta_{ad} \ F^{cb}_{n+m,0}
  - \delta_{cb} \ F^{ad}_{n+m,0} \\
  &   & + \sum_{i=1}^n ( F^{cb}_{n+m-i,0} F^{ad}_{i-1,0}
                         -  F^{cb}_{n-i,0} F^{ad}_{m+i-1,0} )
        + \sum_{i=1}^m ( F^{cb}_{n+i-1,0} F^{ad}_{m-i,0}
                         -  F^{cb}_{i-1,0} F^{ad}_{m+n-i,0} ). \nonumber
\eea
It is also important to note, as in the case of rational potentials, the
existence of $N$ commuting Hamiltonians with an external field, defined as:
\beq
  H^{(n)} = \frac{1}{2n} \Tr(L^+ L^-)^n  \hspace{1cm} {\rm with} \
  L^{\pm} = L \pm a K \pm b K^{-1}
\eeq
since the Poisson structure for $(L^+ L^-)$ has the same characteristic form
as in the rational case, which was generically defined in \cite{A}:
\bea
  \{ L^+ L^- \poi L^+ L^- \} = &   & [ \cR , L^+ L^- \otimes \un ]
  - [ \cRR , \un \otimes L^+ L^- ]  \nonumber \\
  & - & 2 a \ \CC \ ( L^+ L^- \otimes K - K \otimes L^+ L^-
                     + L^+ K L^- \otimes \un - \un \otimes L^+ K L^- ) \\
  & + & 2 b \ \CC \ ( L^+ L^- \otimes K^{-1} - K^{-1} \otimes L^+ L^-
                     + L^+ K^{-1} L^- \otimes \un - \un \otimes L^+ K^{-1}
                     L^- ) \nonumber
\eea
with
\beq
  \cR = (\un \otimes L^+) \ R + R (\un \otimes L^-).
\eeq
This relation is known to ensure commutation of the traces $\Tr(L^+ L^-)^n$
\cite{A}. Such traces are naturally (polynomial) functions of the observables
$\Tr(L^n K^m)$ and $\Tr(L^n K^m F^{ab})$, but it is much easier to see their
commutitativity under the $L^{\pm}$ form. The field-extended models are
therefore ``partially'' integrable. They were considered also in \cite{Po}.

Again one can define a $N \rightarrow \infty$ limit for this algebra,
normalizing each generator by the power of $L$ (the power of $K$ is conserved
hence it is irrelevant, which is obvious since $K=\exp(2Q)$ is anyway
dimensionless):
\bea
  W_n^m & = & N^{n-1} \WW_n^m \\
  \FF^{ab}_{n,m} & = & N^n \FFF^{ab}_{n,m}.
\eea
All reordering terms are eliminated; unfortunately the subleading terms
which were contributing to the Yangian structure are also eliminated and the
remaining structure is a simple ${\cal W}_{\infty}$-plus-current algebra.
Interestingly, the ${\cal W}_{\infty}$ part is the algebra $Sdiff(\R
\times {\rm S}^1)$:
\beq
  \{ \WW_{n_1}^{m_1} , \WW_{n_2}^{m_2} \} = 2 ( n_1 m_2 - n_2 m_1 )
  \WW_{n_1+n_2-1}^{m_1+m_2}.
\eeq
However, this algebra can be redefined as $Sdiff(\R^2)$ by the change
of indexation
\beq
  {\cal W}_n^{n+m} = \WW_n^m.
\eeq
In particular the trigonometric Hamiltonian $\WW_2^0=1/N \ \Tr(L^2)$ is
identified with the rational generator ${\cal W}_2^2 \sim \Tr(L^2 Q^2)$. This
identification was mentioned in the large $N$ limit in \cite{HW2}. Similarly
all commuting trigonometric Hamiltonians become (for $N \rightarrow \infty$)
particular rational observables ---an interesting algebraic connection,
though yet restricted to $N \rightarrow \infty$.

\section{Conclusion}
We have explicitly constructed polynomial algebras of observables for the
ECM models, both rational and trigonometric.
Due to their high degree of intricacy, they do not seem to help in directly
solving the model ---as could be hoped from \cite{HW2}--- when $N$ is finite.
They provide us however with a totally unified structure for the commuting
Hamiltonians, the current-algebra or Yangian-symmetry generators, and an
interesting Virasoro algebra of ``intertwiners'' in the rational case. They
also suggest an algebraic relation between the rational and trigonometric
cases though yet only achieved for infinite $N$.

The case of ECM models in external fields is more hopeful. Indeed one expects
as in the case of pure CM models \cite{A,Pe} to extract from the algebra
of observables ---expressed in the suitable $\{ L^+ , L^-\}$ basis---
exact eigenfunctions of the Hamiltonian flows, at least the $N$ commuting
flows yet known.

\bigskip

\bigskip

\noindent{\Large \bf Acknowledgements} $\ $ We wish to thank O. Babelon and
D. Bernard for discussions. J.A. thanks P. Sorba for his hospitality at
LPT, ENS Lyon. Part of this work was done at SISSA Trieste under EEC contract
540081.

\end{document}